\documentclass[12pt,english,manuscript,aps,prd,showpacs]{revtex4}
\usepackage[T1]{fontenc}
\usepackage[latin9]{inputenc}
\usepackage{graphicx}

\makeatletter

\providecommand{\tabularnewline}{\\}

\@ifundefined{textcolor}{}
{%
 \definecolor{BLACK}{gray}{0}
 \definecolor{WHITE}{gray}{1}
 \definecolor{RED}{rgb}{1,0,0}
 \definecolor{GREEN}{rgb}{0,1,0}
 \definecolor{BLUE}{rgb}{0,0,1}
 \definecolor{CYAN}{cmyk}{1,0,0,0}
 \definecolor{MAGENTA}{cmyk}{0,1,0,0}
 \definecolor{YELLOW}{cmyk}{0,0,1,0}
 }

\makeatother

\usepackage{babel}

\makeatother

\usepackage{babel}

\begin{document}

\title{Anomalous single production of fourth family up type quark associated
with neutral gauge bosons at the LHC}

\author{O. Çak\i{}r}

\email{ocakir@science.ankara.edu.tr}

\affiliation{Department of Physics, Ankara University, 06100, Ankara, Turkey}

\author{I.T. Çak\i{}r}

\email{tcakir@mail.cern.ch}

\affiliation{Department of Physics, CERN, 1211, Geneva 23, Switzerland}

\author{A. Senol}

\email{asenol@kastamonu.edu.tr}

\affiliation{Department of Physics, Kastamonu University, 37100, Kastamonu, Turkey}

\author{A.T. Tasci}

\email{atasci@kastamonu.edu.tr}

\affiliation{Department of Physics, Kastamonu University, 37100, Kastamonu, Turkey}
\begin{abstract}
From the present limits on the masses and mixings of fourth family
quarks, they are expected to have mass larger than the top quark and
allow a large range of mixing of the third family. They could also
have different dynamics than the quarks of three families of the Standard
Model. The single production of the fourth family up type quark $t'$
has been studied via anomalous production process $pp\to t'VX$ (where
$V=g,Z,\gamma$) at the LHC with the center of mass energy of $7$
and $14$ TeV. The signatures of such process are discussed within
both the SM decay modes and anomalous decay modes of $t'$ quarks.
The sensitivity to anomalous coupling $\kappa/\Lambda=0.004$ TeV$^{-1}$
can be reached at $\sqrt{s}=14$ TeV and $L_{int}=100$ pb$^{-1}$.
\end{abstract}


\maketitle

\section{INTRODUCTION}

The number of fermion families in the Standard Model (SM) and its
very extensions remain arbitrary. There is a number of experimental
discrepancies with the SM expectations at some level of standard deviations,
and they can also be interpreted as the loop effects or the flavor
changing neutral current (FCNC) effects by the fourth family fermions.
A minimal framework can give rise to FCNC effects through exchange
of massive SM bosons whose couplings to the light fermions get modified
by mixing with the fourth family fermions. It is also important to
interpret the low energy measurements for the contributions from extra
fermions and analyze them in view of the latest electroweak precision
data \cite{He01,Erler10}.

The current limits on the masses of the fourth SM family leptons are
\cite{Nakamura10}: $m_{l'}>100$ GeV, $m_{\nu'}>90$ ($80$) GeV
for Dirac (Majorana) neutrinos. Recently, the Collider Detector at
Fermilab (CDF) has constrained the masses of the fourth SM family
quarks: $m_{t'}>335$ GeV at 95\% CL. \cite{Conway}, $m_{b'}>338$
GeV at 95\% CL. \cite{Aaltonen}. On the other hand, the perturbativity
of the Yukawa coupling implies that $m_{t'}\leq600$ GeV, and the
arguments based on the partial wave unitarity restrict to an upper
bound $m_{t'}\approx500$ GeV \cite{Chanowitz78}. The parameter space
of fourth family fermion masses with minimal contributions to the
oblique parameters $S$ and $T$, and in agreement with all experimental
constraints, leads to a quark mass splitting $m_{t'}-m_{b'}\simeq50+10 \ln (m_{h}/115\mbox{GeV})$ \cite{Kribs07},
where $m_{h}$ is the mass of Higgs boson. The heavy quark mass bounds
may be relaxed if they are considered together with the fourth family
leptons, which could help in the partial cancellation of the effect
of fourth family on the $S$ and $T$ parameters. Even in the assumption
of the absence of this cancellation the precision measurements restrict
$m_{l'}-m_{\nu'}\approx50-100\mbox{ GeV}$. A review of theoretical
and experimental motivations can be found in \cite{Holdom09}.

\textit{\emph{If the fourth family quarks are discovered at the LHC,
it is expected that it plays a crucial role solving some of the currently
open questions such as the CP violation and flavor structure of the
standard theory \cite{Hou:2010,BarShalom:2009sk,Buras:2010pi,Soni:2008bc,Eberhardt:2010bm,Soni:2010xh,Alok:2010zj},
electroweak symmetry breaking \cite{Holdom:1986rn,Hill:1990ge,Elliott:1992xg,Hung:2010xh},
hierarchies of fermion mass spectrum and mixing angle in quark/lepton
sectors \cite{Holdom:2006mr,Hung:2007ak,Hung:2009ia,Hung:2009hy}.}}
The fourth family quark masses and mixings can be predicted in the
allowed parameter space which is in the reach of the LHC. In the four
families scenario, the assumption of the unitarity of $4\times4$
mixing matrix holds and it allows a large range of the elements as
far as the magnitudes and phases of the $V_{tq}$ are concerned.\emph{
}\textit{\emph{A serious contribution can be expected for the production
of the fourth family fermions from the anomalous interactions. These
effects of the fourth family quarks have been studied in detail at
future linear colliders \cite{asenol}, ep colliders \cite{Alan:2003za,Cakir:2009ir},
and recently at hadron colliders \cite{Sahin:2011zz}. The anomalous
resonant production of fourth family up type quarks via $gq_{i}\to t'$
($q_{i}=u,c$) subprocess \cite{Cakir09,Ciftci:2008tc} and single
production of fourth family quarks \cite{Cakir:2008su} have been
studied at the LHC.}}

In this study, we investigate the production of fourth family up type
quark ($t'$) with the association of the gauge bosons ($g,Z,\gamma$)
through anomalous interactions at the LHC. Then, we consider $t'$
decays into SM and anomalous modes for some relevant parameter space.
We discuss the decay modes for some parameter space and the $t'$
signal observability at three different energies $7$ and $14$ TeV.

\section{Interactions with Fourth Family Quarks}

The interaction Lagrangian for the fourth family quarks ($Q'=t',b'$)
within the SM is given by

\begin{eqnarray}
L_{S} & =-g_{e} & \sum_{Q_{i}'=t',b'}Q_{q_{i}}\overline{Q_{i}'}\gamma^{\mu}Q_{i}'A_{\mu}-g_{s}\sum_{Q_{i}'=t',b'}\overline{Q_{i}'}T^{a}\gamma^{\mu}Q_{i}'G_{\mu}^{a}\nonumber \\
 &  & -\frac{g_{z}}{2}\sum_{Q_{i}'=t',b'}\overline{Q_{i}'}\gamma^{\mu}(g_{v}^{i}-g_{a}^{i}\gamma_{5})Q_{i}'Z_{\mu}\nonumber \\
 &  & -\frac{g_{e}}{2\sqrt{2}\sin\theta_{w}}\sum_{Q'_{i}=t',b'}V_{ij}\overline{Q_{i}'}\gamma^{\mu}(1-\gamma_{5})q_{j}W_{\mu}^{\pm}+\mbox{H.c.}\label{eq:1}\end{eqnarray}
 where $A_{\mu}$, $G_{\mu}$, $Z_{\mu}$ and $W_{\mu}^{\pm}$ are
the fields for photon, gluon, $Z$-boson and $W^{\pm}$ bosons, respectively.
$T_{a}=\lambda_{a}/2$ are Gell-Mann matrices; $Q_{q}$ is the charge
of the quark; $g_{e}$, $g_{z}$ and $g_{s}$ are the electromagnetic,
weak neutral-current and strong coupling constants, respectively.
The coupling $g_{z}=g_{e}/\sin\theta_{W}\cos\theta_{W}$ where $\theta_{W}$
is the Weinberg angle. The $g_{v}$ and $g_{a}$ are the vector and
axial-vector couplings of the neutral weak current with fourth family
quarks. The elements of the extended CKM mixing matrix are presented
by $V_{ij}$. The effective Lagrangian for the FCNC anomalous interactions
between the fourth family quarks, ordinary quarks and the gauge bosons
$V$ ($g,Z,\gamma$) can be written as follows

\begin{eqnarray}
L_{A} & = & \sum_{q_{i}=u,c,t}\frac{\kappa_{\gamma}^{q_{i}}}{\Lambda}Q_{q_{i}}g_{e}\overline{t}'\sigma_{\mu\nu}q_{i}F^{\mu\nu}+\sum_{q_{i}=u,c,t}\frac{\kappa_{z}^{q_{i}}}{2\Lambda}g_{z}\overline{t}'\sigma_{\mu\nu}q_{i}Z^{\mu\nu}\nonumber \\
 &  & +\sum_{q_{i}=u,c,t}\frac{\kappa_{g}^{q_{i}}}{\Lambda}g_{s}\overline{t}'\sigma_{\mu\nu}T_{a}q_{i}G_{a}^{\mu\nu}+h.c.\nonumber \\
 &  & +\sum_{q_{i}=d,s,b}\frac{\kappa_{\gamma}^{q_{i}}}{\Lambda}Q_{q_{i}}g_{e}\overline{b}'\sigma_{\mu\nu}q_{i}F^{\mu\nu}+\sum_{q_{i}=d,s,b}\frac{\kappa_{z}^{q_{i}}}{2\Lambda}g_{z}\overline{b}'\sigma_{\mu\nu}q_{i}Z^{\mu\nu}\nonumber \\
 &  & +\sum_{q_{i}=d,s,b}\frac{\kappa_{g}^{q_{i}}}{\Lambda}g_{s}\overline{b}'\sigma_{\mu\nu}T_{a}q_{i}G_{a}^{\mu\nu}+h.c.\label{eq:3}\end{eqnarray}
 where $F_{\mu\nu}$, $Z_{\mu\nu}$, and $G_{\mu\nu}$ are the field
strength tensors of the photon, $Z$ boson and gluons, respectively;
the $\kappa_{\gamma}$, $\kappa_{z}$ and $\kappa_{g}$ define the
strength of the anomalous couplings for the neutral currents with
the photon, $Z$ boson and gluon, respectively. $\Lambda$ is the
scale for new physics. We have implemented both the SM and anomalous
interaction vertices for the fourth family quarks into the CompHEP
\cite{Boos04} and we use parton distribution library CTEQ6M \cite{CTEQ6M}
with the factorization scale $Q=m_{t'}$.

{\small Here, we have used the parametrization in which the values
for the elements $\mid V_{t'd}\mid=0.0044$, $\mid V_{t'b}\mid=0.22$
and $\mid V_{t's}\mid=0.114$ are given by Ref. \cite{Hou:2006mx}.
In order to reduce free parameters we assume that the anomalous couplings
$\kappa_{z}^{q}$, $\kappa_{g}^{q}$ and $\kappa_{\gamma}^{q}$ are
equal to $\kappa$. The analytical expressions for the decay widths
of $t'$ quark both in SM and anomalous modes are given in Ref. \cite{Cakir09}.
The branching ratios into both SM and anomalous decay modes are given
in Table \ref{tab1}. }It is seen from Table \ref{tab1} that the
anomalous decay mode of the $t'\to qg$ is dominant for $\kappa/\Lambda=1$
TeV$^{-1}$.

\begin{table}
\caption{The branching ratios BR($\%$) and total decay widths ($\Gamma$)
of $t'$ quark depending on the mass ($m_{t'}$), here we take the
anomalous coupling $\kappa/\Lambda=0.1(1)$ TeV$^{-1}$. \label{tab1}}

\begin{tabular}{|c|c|c|c|c|c|c|c|c|c|c|}
\hline 
{\tiny $\begin{array}{c}
\mbox{BR}(\%)\hookrightarrow\\
m_{t'}(\mbox{GeV})\downarrow\end{array}$ }  & {\tiny $Wb$ }  & {\tiny $Ws$ }  & {\tiny $Wd$ }  & {\tiny $gu/gc$ }  & {\tiny $gt$ }  & {\tiny $Zu/Zc$ }  & {\tiny $Zt$ }  & {\tiny $\gamma u/\gamma c$ }  & {\tiny $\gamma t$ }  & {\tiny $\Gamma$(GeV)}\tabularnewline
\hline
\hline 
{\tiny $300$ }  & {\tiny 72.00(7.40) }  & {\tiny 19.00(2.00) }  & {\tiny 0.02(0.003) }  & {\tiny 3.60(37.00) }  & {\tiny 1.00(11.00) }  & {\tiny 0.22(2.20) }  & {\tiny 0.02(0.24) }  & {\tiny 0.08(0.82) }  & {\tiny 0.02(0.24) }  & {\tiny 0.59(5.75)}\tabularnewline
\hline 
{\tiny $400$ }  & {\tiny 71.00(6.80) }  & {\tiny 19.00(1.80) }  & {\tiny 0.03(0.003) }  & {\tiny 3.50(33.00) }  & {\tiny 1.90(18.00) }  & {\tiny 0.23(2.20) }  & {\tiny 0.10(0.95) }  & {\tiny 0.08(0.74) }  & {\tiny 0.04(0.39) }  & {\tiny 1.43(15.02)}\tabularnewline
\hline 
{\tiny $500$ }  & {\tiny 71.00(6.40) }  & {\tiny 19.00(1.70) }  & {\tiny 0.03(0.003) }  & {\tiny 3.50(32.00) }  & {\tiny 2.30(21.00) }  & {\tiny 0.23(2.10) }  & {\tiny 0.15(1.30) }  & {\tiny 0.08(0.70) }  & {\tiny 0.05(0.48) }  & {\tiny 2.82(30.98)}\tabularnewline
\hline 
{\tiny $600$ }  & {\tiny 71.00(6.20) }  & {\tiny 19.00(1.70) }  & {\tiny 0.03(0.003) }  & {\tiny 3.50(31.00) }  & {\tiny 2.60(23.00) }  & {\tiny 0.23(2.10) }  & {\tiny 0.17(1.50) }  & {\tiny 0.08(0.68) }  & {\tiny 0.06(0.52) }  & {\tiny 4.89(55.27)}\tabularnewline
\hline 
{\tiny $700$ }  & {\tiny 70.00(6.10) }  & {\tiny 19.00(1.60) }  & {\tiny 0.03(0.003) }  & {\tiny 3.40(30.00) }  & {\tiny 2.80(25.00) }  & {\tiny 0.24(2.10) }  & {\tiny 0.19(1.70) }  & {\tiny 0.08(0.67) }  & {\tiny 0.06(0.55) }  & {\tiny 7.78(89.52)}\tabularnewline
\hline 
{\tiny $800$ }  & {\tiny 70.00(6.00) }  & {\tiny 19.00(1.60) }  & {\tiny 0.03(0.003) }  & {\tiny 3.40(30.00) }  & {\tiny 3.00(26.00) }  & {\tiny 0.24(2.00) }  & {\tiny 0.20(1.70) }  & {\tiny 0.08(0.66) }  & {\tiny 0.07(0.57) }  & {\tiny 11.64(135.40)}\tabularnewline
\hline
\end{tabular}
\end{table}

\begin{figure}
{\small \includegraphics[scale=0.7]{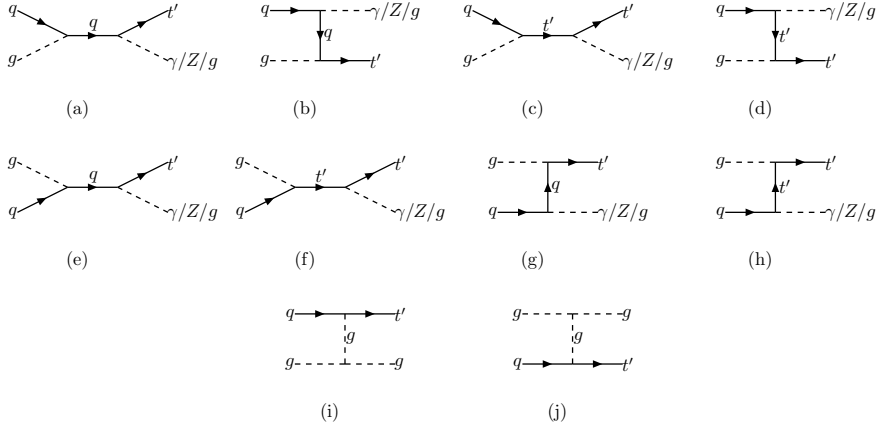} \caption{Feynman diagrams for anomalous $t'V$ ($V=g,Z,\gamma$) production
at the LHC. \label{fig1}}
}
\end{figure}

The Feynman diagrams for the single production of $t'$ quark together
with the vector boson $V$ ($g,Z,\gamma$) are presented in Fig. \ref{fig1}.
Here, the last two diagrams are relevant only for the $t'g$ production
process. The analytical expressions of the differential cross sections
for the subprocess $qg\to t'V$ ($V=g,Z,\gamma$) are calculated and
given in the Appendix.

\section{Numerical Calculations}

\begin{figure}
{\small \includegraphics[scale=0.7]{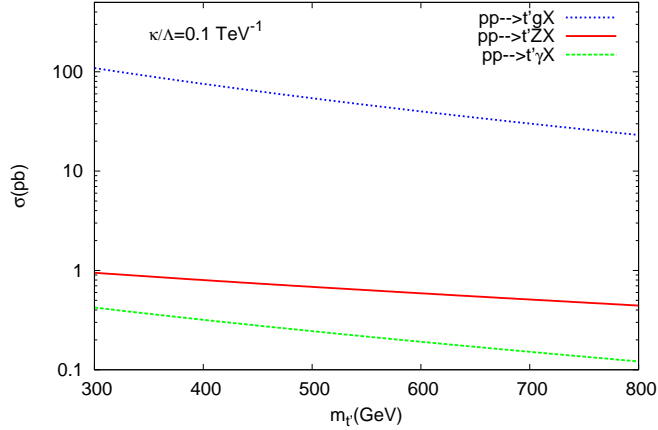} \caption{The cross sections for anomalous $t'V$ ($V=g,Z,\gamma$) production
at the LHC with 14 TeV.{} \label{fig2}}
}
\end{figure}

The total cross sections for the $t'V$ ($V=g,Z,\gamma$) production
with $\kappa/\Lambda=0.1$ TeV$^{-1}$ are presented in Fig. \ref{fig2}
depending on the mass $m_{t'}$ at the collision center of mass energy
of $14$ TeV. As seen from Fig. \ref{fig2} the cross section for
the process $pp\to t'gX$ is much larger than that of the $pp\to t'\gamma X$
and $pp\to t'ZX$ processes. Due to large QCD background in the $qgg$
production (which includes only one type anomalous coupling $\kappa_{g}$
for the case $t'\to W^{+}q$) and small branching ratio into $q\gamma/Z$
for $\kappa/\Lambda\leq0.1$ TeV$^{-1}$ as seen from Table \ref{tab1},
we consider only the processes $pp\to t'ZX$ and $pp\to t'\gamma X$
with subsequent decay into SM mode $t'\to W^{+}b$ for both processes,
and the anomalous mode $t'\to tg$ for the first process. Hereafter,
in the numerical calculations we use the anomalous coupling $\kappa/\Lambda=0.1$
TeV$^{-1}$.

\begin{figure}
{\small \includegraphics[width=8cm]{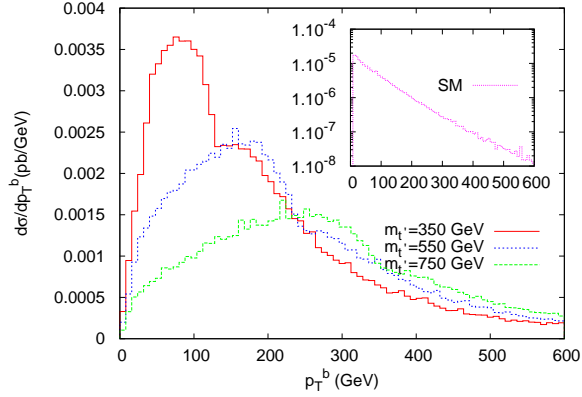} \caption{The $p_{T}$ distributions of $b$ quark from signal and background
process $pp\to W^{+}bZX$ at $\sqrt{s}$=14 TeV. \label{fig3}}
}
\end{figure}

\begin{figure}
{\small \includegraphics[width=8cm]{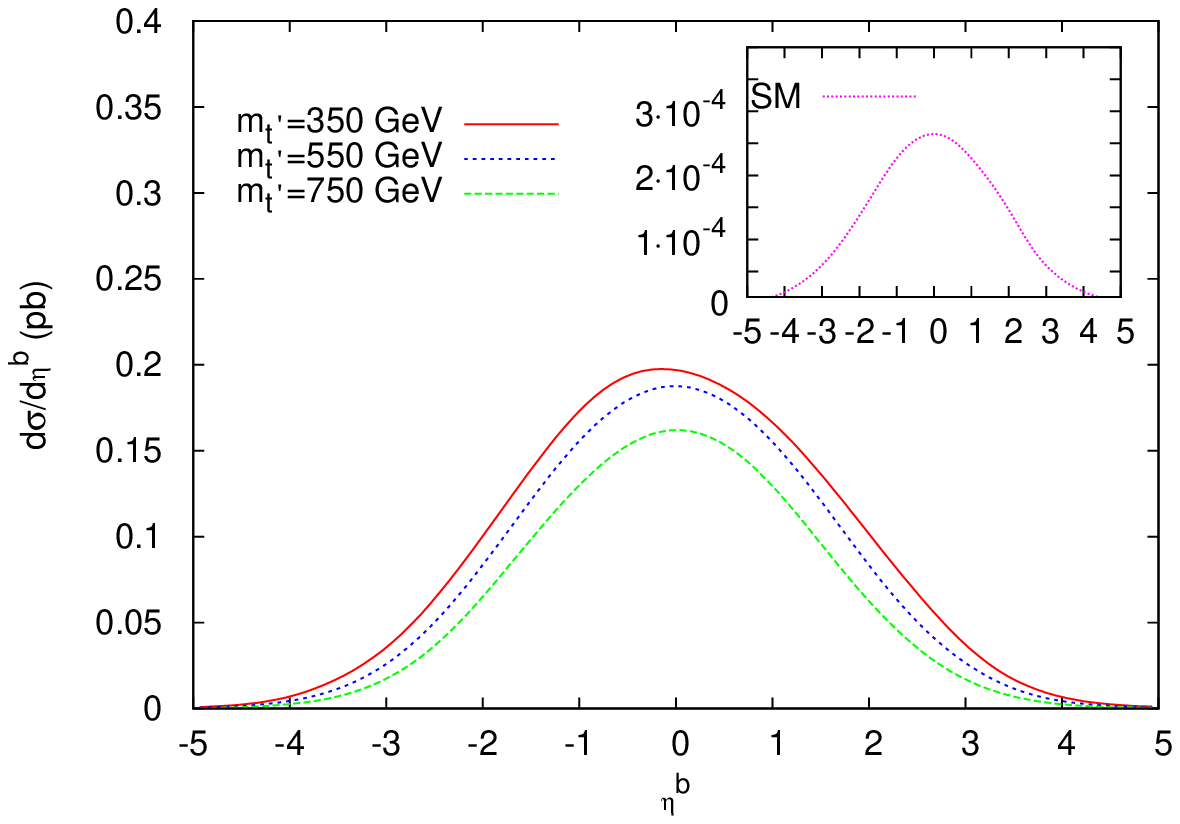} \caption{The $\eta$ distributions of $b$ quark from signal and background
process $pp\to W^{+}bZX$ at $\sqrt{s}$=14 TeV. \label{fig4}}
}
\end{figure}

\begin{figure}
{\small \includegraphics[width=8cm]{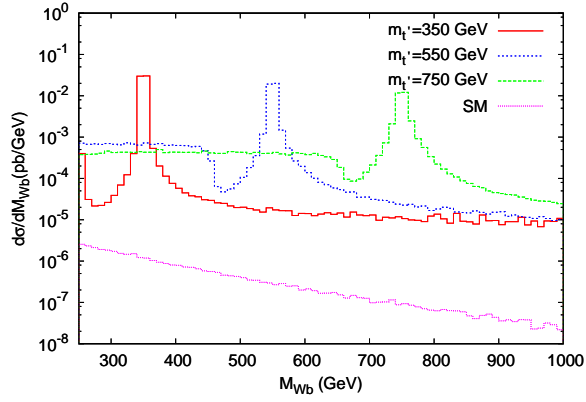} \caption{Invariant mass distributions of the $W^{+}b$ system ($pp\to W^{+}bZX$)
for the given parametrization at $\sqrt{s}$=14 TeV. \label{fig5}}
}
\end{figure}

Transverse momentum distributions of the $b$ quark for the signal
$pp\to W^{+}bZX$ (where $m_{t'}=$350, 550, 750 GeV) and the SM background
are shown in Fig. \ref{fig3}. The signal differential cross sections
show peak around $p_{T}=75$, 150 and 250 GeV for the values of $m_{t'}=350$,
550 and 750 GeV, respectively. The background differential cross section
smoothly decreases with respect to the $p_{T}$ of $b$ quark.{\small{}
The rapidity distributions of the $b$ quark from the signal and background
locate mostly in the range of $|\eta^{b}|<2$ as shown in Fig. \ref{fig4}.
}The signal has peak around the $t'$ mass values of 350, 550 and
750 GeV over the smooth background. The level of the relevant background
is compared with the signal in the invariant mass distributions of
$W^{+}b$ system as shown in Fig. \ref{fig5}. 

\begin{figure}
{\small \includegraphics[width=8cm]{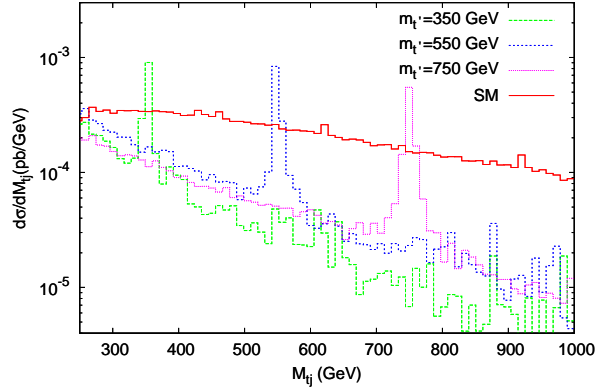} \caption{Invariant mass distributions of the $t+jet$ system ($pp\to tgZX$
for signal and $pp\to tjZX$ for background) at $\sqrt{s}$=14 TeV.
\label{fig6}}
}
\end{figure}

The differential cross sections for the signal process $pp\to tgZX$
in the $tg$ invariant mass distributions are given in Fig. \ref{fig6}.
The signal shows itself as the peaks around the assumed mass values
of $m_{t'}$. Here, we take into account the process $pp\to tjZX$
for background where $j$ denotes light quarks in the final state.

\begin{figure}
{\small \includegraphics[width=10cm]{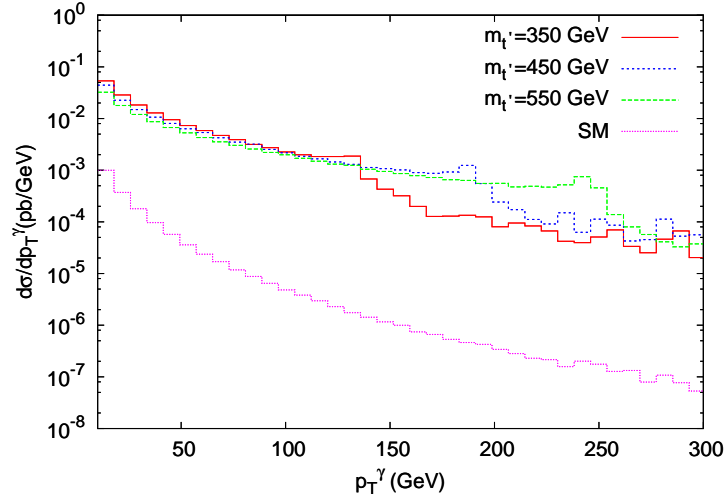} \caption{The $p_{T}$ distributions of photon from the signal and background
at $\sqrt{s}=14$ TeV.\label{fig7}}
}
\end{figure}

The $p_{T}$ distributions of the photon from the signal and background
processes $pp\to W^{+}b\gamma X$ are shown in Fig. \ref{fig7}.

In order to calculate the statistical significance ($SS$) we use
the following formula

\[
SS=\sqrt{2[(S+B)\ln(1+S/B)-S]}\]
 where $S$ and $B$ are the number of events of the signal and background,
respectively. In the analysis, we use the kinematical cuts on the
transverse momentum $p_{T}^{j}>50$ GeV of jets and $p_{T}^{\gamma}>20$
GeV of photons. We calculate the signal and background cross sections
for the process $pp\to W^{+}b(Z,\gamma)X$ in the invariant mass interval
satisfying the conditions $|m_{t'}-M_{Wb}|<10$ GeV for the $t'$
mass range 300-500 GeV and $|m_{t'}-M_{Wb}|<20$ GeV for the $t'$
mass range 500-800 GeV. In the significance calculations for the signal
process $pp\to W^{+}b(Z,\gamma)X$, we take into account the hadronic
$W$- decay and leptonic $Z$-decay. We use $b$-tagging efficiency
as 0.5 for considered processes.

The $SS$ values, total cross sections of the signal and background
process $pp\to W^{+}bZ(\gamma)X$ are given in Table \ref{tab:table2}(\ref{tab:table4})
for $\sqrt{s}$=14 TeV and $L=10^{5}\:\mbox{pb}{}^{-1}$, and in Table
\ref{tab:table3}(\ref{tab:table5}) for $\sqrt{s}$=7 TeV and $L=10^{3}\:\mbox{pb}{}^{-1}$,
respectively.

\begin{table}
{\small \caption{The total cross sections of the signal and background process $pp\to W^{+}bZX$,
as well as the $SS$ values at $\sqrt{s}$=14 TeV with $L=10^{5}\mbox{pb}{}^{-1}$.\label{tab:table2}}
}{\small \par}

\begin{tabular}{lcccccl}
\hline 
{\small $m_{t'}(\mbox{GeV})$ }  &  & {\small $10^{-5}\times\sigma_{B}$(pb)}  &  & {\small $10^{-1}\times\sigma_{S}$(pb) }  &  & {\small $SS$ }\tabularnewline
\hline 
{\small 300 }  &  & {\small 4.85 }  &  & {\small 6.70 }  &  & {\small 165 }\tabularnewline
{\small 400 }  &  & {\small 2.21 }  &  & {\small 5.42 }  &  & {\small 153 }\tabularnewline
{\small 500 }  &  & {\small 1.05 }  &  & {\small 4.38 }  &  & {\small 142 }\tabularnewline
{\small 600 }  &  & {\small 1.10 }  &  & {\small 3.82 }  &  & {\small 131 }\tabularnewline
{\small 700 }  &  & {\small 0.62 }  &  & {\small 3.15 }  &  & {\small 121 }\tabularnewline
{\small 800 }  &  & {\small 0.37 }  &  & {\small 2.56 }  &  & {\small 111 }\tabularnewline
\hline
\end{tabular}
\end{table}

\begin{table}
{\small \caption{The same as Table \ref{tab:table2}, but for $\sqrt{s}$=7 TeV and
$L=10^{3}$pb$^{-1}$.\label{tab:table3}}
}{\small \par}

\begin{tabular}{lcccccl}
\hline 
{\small $m_{t'}$(GeV) }  &  & {\small $10^{-6}\times\sigma_{B}$(pb) }  &  & {\small $10^{-1}\times\sigma_{S}$(pb) }  &  & {\small $SS$ }\tabularnewline
\hline 
{\small 300 }  &  & {\small 8.54 }  &  & {\small 1.23 }  &  & {\small 7.1 }\tabularnewline
{\small 400 }  &  & {\small 3.56 }  &  & {\small 0.86 }  &  & {\small 6.1 }\tabularnewline
{\small 500 }  &  & {\small 1.56 }  &  & {\small 0.62 }  &  & {\small 5.3 }\tabularnewline
{\small 600 }  &  & {\small 1.50 }  &  & {\small 0.47 }  &  & {\small 4.6 }\tabularnewline
{\small 700 }  &  & {\small 0.77 }  &  & {\small 0.34 }  &  & {\small 4.0 }\tabularnewline
{\small 800 }  &  & {\small 0.41 }  &  & {\small 0.25 }  &  & {\small 3.4 }\tabularnewline
\hline
\end{tabular}
\end{table}

\begin{table}
{\small \caption{The $SS$ values, total cross sections of the signal and background
process $pp\to t\gamma ZX$ at $\sqrt{s}$=14 TeV with $L=10^{5}\mbox{pb}{}^{-1}$.
\label{tab:table4}}
}{\small \par}

\begin{tabular}{lcccl}
\hline 
{\small $m_{t'}$(GeV) }  &  & {\small $10^{-6}\times\sigma_{B}$(pb) }  & {\small $10^{-1}\times\sigma_{S}$(pb) }  & {\small $SS$ }\tabularnewline
\hline 
{\small 300 }  &  & {\small 24.6 }  & {\small 2.68 }  & {\small 389 }\tabularnewline
{\small 400 }  &  & {\small 9.87 }  & {\small 3.12 }  & {\small 446 }\tabularnewline
{\small 500 }  &  & {\small 4.43 }  & {\small 3.03 }  & {\small 457 }\tabularnewline
{\small 600 }  &  & {\small 4.36 }  & {\small 2.75 }  & {\small 434 }\tabularnewline
{\small 700 }  &  & {\small 2.26 }  & {\small 2.51 }  & {\small 426 }\tabularnewline
{\small 800 }  &  & {\small 1.25 }  & {\small 2.28 }  & {\small 415 }\tabularnewline
\hline
\end{tabular}
\end{table}

\begin{table}
{\small \caption{The same as Table \ref{tab:table4}, but for $\sqrt{s}$=7 TeV and
$L=10^{3}$ pb$^{-1}$. \label{tab:table5}}
}{\small \par}

\begin{tabular}{lcccl}
\hline 
{\small $m_{t'}$(GeV) }  &  & {\small $10^{-7}\times\sigma_{B}$(pb) }  & {\small $10^{-2}\times\sigma_{S}$(pb) }  & {\small $SS$ }\tabularnewline
\hline 
{\small 300 }  &  & {\small 43.5 }  & {\small 7.45 }  & {\small 21.1 }\tabularnewline
{\small 400 }  &  & {\small 15.5 }  & {\small 8.30 }  & {\small 23.6 }\tabularnewline
{\small 500 }  &  & {\small 6.18 }  & {\small 7.42 }  & {\small 23.2 }\tabularnewline
{\small 600 }  &  & {\small 5.17 }  & {\small 6.13 }  & {\small 21.1 }\tabularnewline
{\small 700 }  &  & {\small 2.44 }  & {\small 5.05 }  & {\small 19.7 }\tabularnewline
{\small 800 }  &  & {\small 1.19 }  & {\small 4.03 }  & {\small 17.9 }\tabularnewline
\hline
\end{tabular}
\end{table}

In Tables \ref{tab:table6} and \ref{tab:table7}, we consider leptonic
(hadronic) decays of $Z$-boson and hadronic (leptonic) decays of
$W$-boson for the process $pp\to tjZX$. The conditions $|m_{t'}-M_{tj}|<10$
GeV for $m_{t'}=$300-500 GeV and $|m_{t'}-M_{tj}|<20$ GeV for $m_{t'}=$500-800
GeV are used for the process $pp\to tjZX$. Taking the integrated
luminosity $L=10^{5}\:\mbox{pb}{}^{-1}$, the $SS$ values and the
total cross sections of the signal and background are given in Table
\ref{tab:table6} for $\sqrt{s}$=14 TeV.

\begin{table}
{\small \caption{The total cross sections of the signal and background process $pp\to tgZX$
and $SS$ values which corresponds to the leptonic (hadronic) decays
of $Z$-boson and hadronic (leptonic) decays of $W$-boson at $\sqrt{s}$=14
TeV with $L=10^{5}\mbox{pb}{}^{-1}$. \label{tab:table6}}
}{\small \par}

\begin{tabular}{lccc}
\hline 
{\small $m_{t'}$(GeV) }  & {\small $10^{-3}\times\sigma_{B}$(pb) }  & {\small $10^{-3}\times\sigma_{S}$(pb) }  & {\small $SS$}\tabularnewline
\hline 
{\small 300 }  & {\small 0.75 }  & {\small 7.94 }  & {\small 8.0(14.3) }\tabularnewline
{\small 350 }  & {\small 1.43 }  & {\small 11.0 }  & {\small 8.7(15.7) }\tabularnewline
{\small 400 }  & {\small 2.19 }  & {\small 12.3 }  & {\small 8.5(15.2) }\tabularnewline
{\small 500 }  & {\small 3.85 }  & {\small 12.2 }  & {\small 7.1(12.8) }\tabularnewline
{\small 600 }  & {\small 5.46 }  & {\small 12.2 }  & {\small 6.4(11.5) }\tabularnewline
{\small 700 }  & {\small 6.94 }  & {\small 10.4 }  & {\small 5.1(9.2) }\tabularnewline
{\small 800 }  & {\small 8.18 }  & {\small 8.60 }  & {\small 4.1(7.3) }\tabularnewline
\hline
\end{tabular}
\end{table}

\begin{table}
{\small \caption{The same as Table \ref{tab:table6}, but for $\sqrt{s}$=7 TeV. \label{tab:table7}}
}{\small \par}

\begin{tabular}{lccl}
\hline 
{\small $m_{t'}$(GeV) }  & {\small $10^{-4}\times\sigma_{B}$(pb) }  & {\small $10^{-3}\times\sigma_{S}$(pb) }  & {\small $SS$ }\tabularnewline
\hline 
{\small 300 }  & {\small 1.82 }  & {\small 1.52 }  & {\small 3.3(5.9) }\tabularnewline
{\small 350 }  & {\small 3.20 }  & {\small 2.00 }  & {\small 3.5(6.3) }\tabularnewline
{\small 400 }  & {\small 4.59 }  & {\small 2.06 }  & {\small 3.3(5.9)}\tabularnewline
{\small 500 }  & {\small 7.14 }  & {\small 1.76 }  & 2.5(4.5)\tabularnewline
{\small 600 }  & {\small 8.96 }  & {\small 1.55 }  & 2.1(3.7)\tabularnewline
{\small 700 }  & {\small 10.1 }  & {\small 1.15 }  & 1.5(2.8)\tabularnewline
{\small 800 }  & {\small 10.6 }  & {\small 0.84 }  & 1.1(2.0)\tabularnewline
\hline
\end{tabular}
\end{table}

\begin{figure}
{\small \includegraphics[width=10cm]{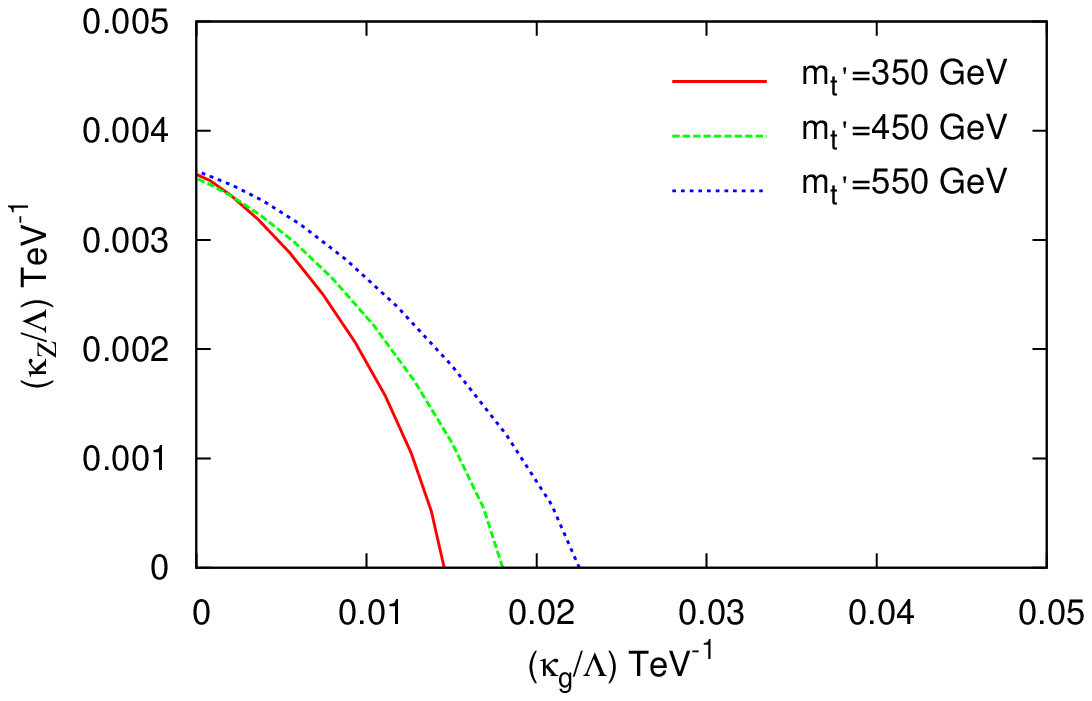} \caption{The $3\sigma$ contour plot for the signal observability at $\sqrt{s}=14$
TeV and $L=10^{5}$ pb$^{-1}$.\label{fig8}}
}
\end{figure}

In Fig. \ref{fig8}, we plot the contours for different mass values,
namely $m_{t'}=$350, 450 and 550 GeV, with the $3\sigma$ significance
for process $pp\to W^{+}bZX$. We find that the lower limit on the
anomalous couplings $\kappa_{g}/\Lambda$ and $\kappa_{z}/\Lambda$
can be as low as 0.015 and 0.0035 TeV$^{-1}$, respectively.

\begin{figure}
{\small \includegraphics[width=10cm]{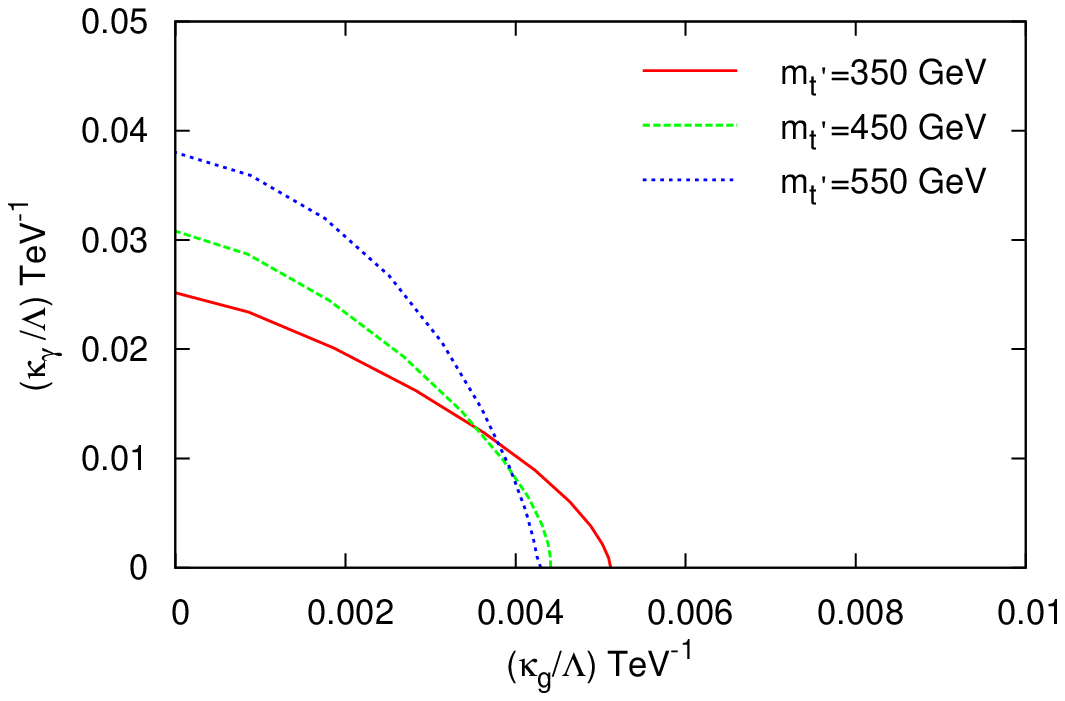} \caption{The same as Fig. \ref{fig8}, but for the anomalous couplings $\kappa_{\gamma}/\Lambda$
vs $\kappa_{g}/\Lambda$.\label{fig9}}
}{\small \par}

\end{figure}

In Fig. \ref{fig9}, we plot the contours for the mass values $m_{t'}=350$,
450 and 550 GeV for $3\sigma$ significance. It is found that lower
limit on the anomalous couplings $\kappa_{g}/\Lambda$ and $\kappa_{\gamma}/\Lambda$
from the process $pp\to W^{+}b\gamma X$ can be as low as 0.004 and
0.025 TeV$^{-1}$, respectively.

\section{Conclusions}

The anomalous FCNC interactions could be important for some parameter
regions due to the expected large masses of the fourth family quarks.
From the analysis of single production of $t'$ quark associated with
the neutral gauge bosons the sensitivity to the ($\kappa_{g},\kappa_{z}$)
and ($\kappa_{g},\kappa_{\gamma}$) can be obtained as low as (0.015,
0.0035 ) and (0.004, 0.025) at the LHC with the center of mass energy
of 14 TeV and the integrated luminosity of $10^{5}$pb$^{-1}$.

{\small \newpage{} }{\small \par}

\section*{APPENDIX}

The differential cross section for the subprocess $qg\to t'\gamma$
is given as

{\small \begin{eqnarray}
\frac{d\hat{\sigma}}{d\hat{t}}(qg\rightarrow t'\gamma) & = & \frac{2\pi\alpha_{s}\alpha}{27\hat{s}^{2}\Lambda^{2}}\Bigg\{\frac{\kappa_{g}^{2}2\hat{s}[m^{2}(m^{2}-6\hat{s})+\hat{s}^{2}]}{[(\hat{s}-m^{2})^{2}+\Gamma^{2}m^{2}]}+\frac{2\kappa_{\gamma}^{2}(m^{2}-\hat{s})^{2}}{\hat{s}}+\kappa_{g}\kappa_{\gamma}\Big[-(\hat{s}+m^{2})\nonumber \\
 & + & \frac{3(2\hat{s}^{3}-m^{6})-m^{2}[2\hat{s}(\hat{s}-8m^{2})+\hat{t}(3m^{2}-\hat{s})])}{[(\hat{s}-m^{2})^{2}+\Gamma^{2}m^{2}]}\Big]+2\Bigg[\Big[\frac{\kappa_{\gamma}^{2}\hat{t}[m^{2}(\hat{u}-2\hat{t})-\hat{t}\hat{u}]}{(\hat{t}-m^{2})^{2}}\nonumber \\
 & + & \frac{\kappa_{g}^{2}\hat{u}(m^{2}-\hat{t})}{\hat{t}}-\frac{\kappa_{g}\kappa_{\gamma}}{4}\frac{[3m^{4}+(3\hat{u}-7\hat{t})m^{2}+\hat{t}(4\hat{t}-\hat{u})]}{\hat{t}-m^{2}}\Big]+[\hat{t}\leftrightarrow\hat{u}]\Bigg]\nonumber \\
 & + & \Big[\frac{\kappa_{\gamma}^{2}}{\hat{s}(\hat{t}-m^{2})(\hat{u}-m^{2})}+\frac{\kappa_{g}^{2}\hat{s}}{\hat{u}\hat{t}[(\hat{s}-m^{2})^{2}+\Gamma^{2}m^{2}]}\Big][m^{2}(\hat{s}-m^{2})[m^{2}(\hat{s}-m^{2})+2\hat{u}\hat{t}]]\nonumber \\
 & + & \frac{\kappa_{\gamma}\kappa_{g}(\hat{s}-m^{2})}{[(\hat{s}-m^{2})^{2}+\Gamma^{2}m^{2}]}\Big[\frac{3m^{4}(\hat{u}-m^{2})+3(\hat{s}^{2}+3\hat{s}\hat{t}+\hat{t}^{2})m^{2}+\hat{s}\hat{t}\hat{u}}{(\hat{t}-m^{2})}\nonumber \\
 & + & \frac{3m^{4}(\hat{t}-\hat{s})-3(\hat{s}^{2}+\hat{s}\hat{t}-\hat{t}^{2})m^{2}+\hat{s}\hat{t}\hat{u}}{(\hat{u}-m^{2})}\Big]\Bigg\}\end{eqnarray}
 }where $\hat{s}$, $\hat{t}$ and $\hat{u}$ are the Mandelstam variables
for the subprocess; $m$ and $\Gamma$ denote the mass and decay width
of $t'$ quark, respectively. The strong and electromagnetic coupling
constants are presented by $\alpha_{s}$ and $\alpha$, respectively.
The differential cross section for the subprocess $qg\to t'g$ is
given as

{\small \begin{eqnarray}
\frac{d\hat{\sigma}}{d\hat{t}}(qg\rightarrow t'g) & = & \frac{\pi\alpha_{s}^{2}\kappa_{g}^{2}}{72\hat{s}^{2}\Lambda^{2}}\Bigg\{\frac{32\hat{s}[m^{2}(m^{2}-6\hat{s})+\hat{s}^{2}]}{[(\hat{s}-m^{2})^{2}+\Gamma^{2}m^{2}]}+\frac{32(m^{2}-\hat{s})^{2}}{\hat{s}}+2(m^{2}+\hat{s})\nonumber \\
 & + & \frac{144(m^{2}-\hat{s})^{3}}{[(\hat{s}-m^{2})^{2}+\Gamma^{2}m^{2}]}+\frac{9[16m^{6}+18m^{4}(\hat{s}-m^{2})+2m^{2}(\hat{u}^{2}+\hat{t}^{2})+3\hat{u}\hat{t}(5m^{2}+\hat{s})]}{\hat{t}\hat{u}}\nonumber \\
 & + & 4\Bigg[\Big[\frac{8[m^{2}(\hat{u}-2\hat{t})-\hat{t}\hat{u}]}{(\hat{t}-m^{2})^{2}}+\frac{9[4\hat{t}\hat{s}+\hat{u}(3m^{2}+\hat{u})-4m^{4}]+8(m^{2}-\hat{t})\hat{u}}{\hat{t}}\nonumber \\
 & + & \frac{9[8m^{6}-6(2\hat{s}+\hat{u})m^{4}+4\hat{s}^{2}(m^{2}-\hat{u})+9\hat{s}\hat{u}m^{2}-\hat{u}^{2}(2m^{2}-\hat{s})]}{4\hat{s}\hat{u}}\nonumber \\
 & + & \frac{8[3m^{4}+(3\hat{u}-7\hat{t})m^{2}+\hat{t}(4\hat{t}-\hat{u})]}{\hat{t}-m^{2}}+\frac{4m^{2}}{\hat{s}}[\frac{(m^{2}-\hat{s})\hat{t}}{\hat{t}-m^{2}}+\frac{\hat{s}^{2}(m^{2}-\hat{u})(\hat{s}-m^{2})}{\hat{t}[(\hat{s}-m^{2})^{2}+\Gamma^{2}m^{2}]}]\nonumber \\
 & + & \frac{[3(m^{2}-\hat{u})m^{4}+(\hat{t}-3m^{2})\hat{s}^{2}+(\hat{s}-3m^{2})\hat{t}^{2}-10\hat{t}\hat{s}m^{2}](\hat{s}-m^{2})}{(\hat{t}-m^{2})[(\hat{s}-m^{2})^{2}+\Gamma^{2}m^{2}]}\Big]+[\hat{t}\leftrightarrow\hat{u}]\Bigg]\\
 & + & \Bigg[[\frac{9[(5\hat{t}-4\hat{s})m^{4}-4\hat{s}^{2}(\hat{t}+m^{2})-3\hat{t}\hat{s}m^{2}+\hat{t}^{2}(\hat{s}-5m^{2})](\hat{s}-m^{2})}{\hat{t}[(\hat{s}-m^{2})^{2}+\Gamma^{2}m^{2}]}]+[\hat{s}\leftrightarrow\hat{u}]\Bigg]\nonumber \\
 & + & \Bigg[[\frac{9[(5\hat{t}-\hat{s})m^{4}+6\hat{s}^{2}(\hat{t}-2m^{2})+t^{2}(\hat{s}-5m^{2})+\hat{s}(5\hat{s}^{2}-9\hat{t}m^{2})](\hat{s}-m^{2})}{\hat{u}[(\hat{s}-m^{2})^{2}+\Gamma^{2}m^{2}]}]\nonumber \\
 & + & [\hat{s}\leftrightarrow\hat{u}]\Bigg]\Bigg\}\nonumber \end{eqnarray}
 }{\small \par}

The differential cross section for the subprocess $qg\to t'Z$ is
given as{\small{} \begin{eqnarray}
\frac{d\hat{\sigma}}{d\hat{t}}(qg\rightarrow t'Z) & = & \frac{\pi\alpha_{s}\alpha}{432\hat{s}^{2}\Lambda^{2}s_{W}^{2}c_{W}^{2}}\Bigg\{\frac{\kappa_{g}^{2}\hat{s}}{m_{Z}^{2}[(\hat{s}-m^{2})^{2}+\Gamma^{2}m^{2}]}[18(g_{v}^{2}+g_{a}^{2})(m^{6}+m^{4}(m_{Z}^{2}-\hat{s})\nonumber \\
 & - & 2m_{Z}^{4}(\hat{s}+m^{2})+\hat{s}^{2}(m_{Z}^{2}-m^{2}+\hat{s})+2\hat{s}m^{2}m_{Z}^{2}(-160s_{W}^{4}+120s_{W}^{2}+9)]\nonumber \\
 & + & \frac{36\kappa_{z}^{2}}{\hat{s}}[2m^{4}-m_{Z}^{2}(\hat{s}+m^{2}+m_{Z}^{2})+2\hat{s}(\hat{s}-2m^{2})]\nonumber \\
 & + & \frac{324g_{v}\kappa_{z}\kappa_{g}(\hat{s}-m^{2})}{[(\hat{s}-m^{2})^{2}+\Gamma^{2}m^{2}]}[m_{Z}^{2}(\hat{s}+m^{2})-\hat{s}(\hat{s}-2m^{2})-m^{4}]\nonumber \\
 & + & 36\Bigg[\Big[\frac{\kappa_{z}^{2}}{(\hat{t}-m^{2})^{2}}[2m^{4}(\hat{t}-m_{Z}^{2})+m^{2}(2m_{Z}^{4}+(\hat{s}+6\hat{t})m_{Z}^{2}-2\hat{t}(\hat{s}+4\hat{t}))\nonumber \\
 & + & \hat{t}(2\hat{t}(\hat{s}+\hat{t})-m_{Z}^{2}(\hat{s}+2\hat{t}))]+\frac{g_{v}\kappa_{g}\kappa_{z}}{\hat{t}}[2m_{Z}^{2}(m^{2}-\hat{t})+\hat{t}(m^{2}-\hat{u})]\nonumber \\
 & + & \frac{\kappa_{g}^{2}(m^{2}-\hat{t})(g_{v}^{2}+g_{a}^{2})}{\hat{t}^{2}}[m^{2}(\hat{t}^{2}-2m_{Z}^{4})-\hat{t}(\hat{t}(\hat{t}+\hat{u})-m_{Z}^{2}(\hat{t}+2\hat{u}))]\nonumber \\
 & + & \frac{2\kappa_{g}^{2}\hat{s}(m^{2}-\hat{s})}{9m_{Z}^{2}\hat{t}[(\hat{s}-m^{2})^{2}+\Gamma^{2}m^{2}]}[9(g_{v}^{2}+g_{a}^{2})((\hat{t}-m^{2})(2m_{Z}^{2}-m^{2}\hat{t})-\hat{s}\hat{t}m^{2})\nonumber \\
 & + & 4s_{W}^{2}\hat{s}(4s_{W}^{2}-3)\hat{t}^{2}]+\frac{2\kappa_{z}^{2}}{\hat{s}(\hat{t}-m^{2})}[m^{2}(\hat{s}-m^{2})(2\hat{t}-m_{Z}^{2})+m_{Z}^{2}(m^{2}(\hat{t}^{2}+m^{2})\nonumber \\
 & + & \hat{t}\hat{s}-\hat{u}m_{Z}^{2})]+\frac{\kappa_{z}\kappa_{g}g_{v}}{\hat{t}(\hat{t}-m^{2})}[(m_{Z}^{2}-\hat{t})(6m^{4}-11\hat{t}m^{2}+5\hat{t}^{2})+\hat{s}\hat{t}(3m^{2}-\hat{t})]\\
 & + & \frac{\kappa_{z}\kappa_{g}g_{v}(\hat{s}-m^{2})}{(\hat{t}-m^{2})[(\hat{s}-m^{2})^{2}+\Gamma^{2}m^{2}]}[3m^{4}(\hat{u}-m^{2})-\hat{s}(\hat{s}(3m^{2}-\hat{t})+3m_{Z}^{2}(\hat{t}-m^{2}))\nonumber \\
 & - & \hat{t}(\hat{t}(3m^{2}-\hat{s})+m^{2}(10\hat{s}-3m_{Z}^{2}))]\Big]+[\hat{t}\leftrightarrow\hat{u}]\Bigg]\Bigg\}\nonumber \end{eqnarray}
 }where $g_{v}$ and $g_{a}$ are the vector and axial-vector couplings
of the weak neutral current; $m_{Z}$ is the mass of the $Z$ boson.

\end{document}